\documentclass[zpreprint,zbstpl]{./zeus_paper}
\usepackage[english]{babel}
\chardef\usc=95
\chardef\til=126
\catcode`\@=11 
\DeclareRobustCommand\xdotspace{\futurelet\@let@token\@xdotspace}
\def\@xdotspace{%
  \ifx\@let@token.\else
  \ifx\@let@token\bgroup.\else
  \ifx\@let@token\egroup.\else
  \ifx\@let@token\/.\else
  \ifx\@let@token\ .\else
  \ifx\@let@token~.\else
  \ifx\@let@token!.\else
  \ifx\@let@token,.\else
  \ifx\@let@token:.\else
  \ifx\@let@token;.\else
  \ifx\@let@token?.\else
  \ifx\@let@token/.\else
  \ifx\@let@token'.\else
  \ifx\@let@token).\else
  \ifx\@let@token-.\else
  \ifx\@let@token\@xobeysp.\else
  \ifx\@let@token\space.\else
  \ifx\@let@token\@sptoken.\else
   .\space
   \fi\fi\fi\fi\fi\fi\fi\fi\fi\fi\fi\fi\fi\fi\fi\fi\fi\fi}
\catcode`\@=12 

\newcommand{\stru}[2]{%
   \relax\ifmmode\hbox{\vrule height#1 depth#2 width0pt}%
   \else\vrule height#1 depth#2 width0pt\fi}

\newcommand{\Ronum}[1]{\uppercase\expandafter{\romannumeral#1}}
\newcommand{\ronum}[1]{\expandafter{\romannumeral#1}}
\DeclareRobustCommand{\LaTeXZ}{%
  \LaTeX\kern-.05em4\kern-.1em
  {\raisebox{-0.2ex}{$\scriptstyle\text{ZEUS}$}}\xspace}

\DeclareMathAlphabet{\mathbf}{OT1}{cmr}{bx}{sl}
\newcommand{\eVdist}{\kern-0.06667em}

\newcommand{\pb}{\,\text{pb}}

\newcommand{\slashfrac}[2]{%
  \raisebox{0.5ex}{\ensuremath #1}\kern-0.12em/\kern-0.08em
  \raisebox{-.8ex}{\ensuremath #2}}

\newcommand{\sqr}[3]{%
    {\vcenter{\hrule height.#3ex\hbox{\vrule width.#2ex height#1ex
     \kern#1ex\vrule width.#3ex}\hrule height.#2ex}}}

\catcode`\@=11 
\newcommand{\parenbar}{\mathpalette\p@renb@r}
\def\p@renb@r#1#2{\vbox{%
  \ifx#1\scriptscriptstyle \dimen@.7em\dimen@ii.2em\else
  \ifx#1\scriptstyle \dimen@.8em\dimen@ii.25em\else
  \dimen@1em\dimen@ii.4em\fi\fi \offinterlineskip
  \ialign{\hfill##\hfill\cr
    \vbox{\hrule width\dimen@ii}\cr
    \noalign{\vskip-.3ex}%
    \hbox to\dimen@{$\mathchar300\hfil\mathchar301$}\cr
    \noalign{\vskip-.3ex}%
    $#1#2$\cr}}}
\catcode`\@=12 

\newcommand{\IP}{{\rm I$\kern-0.01667em$P}\xspace}

\mathchardef\qsm=63
\mathchardef\pls=43
\mathchardef\mns=512
\mathchardef\plm=518
\mathchardef\eql=61
\mathchardef\smallleft=300
\mathchardef\smallright=301
\mathchardef\les=316
\mathchardef\gre=318
\mathchardef\leq=532
\mathchardef\grq=533

\catcode`\@=11 
\newcounter{pict@width}
\newcounter{pict@height}
\newlength{\pict@scale}
\newcommand{\psfigadd}[4]{%
\setcounter{pict@width}{1*\ratio{#2+\pict@scale/2}{\pict@scale}}
\setcounter{pict@height}{1*\ratio{#3+\pict@scale/2}{\pict@scale}}
\setlength{\unitlength}{\pict@scale}
\hbox to #2{\hspace{-\fill}\begin{picture}(\thepict@width,\thepict@height)
\put(0,0){\psfig{figure=#1,width=#2,height=#3,clip=}}
\SetScale{0.283466457}
\SetWidth{1.763889}
{#4}
\end{picture}}
}
\newcounter{pict@widthfst}
\newcounter{pict@widthscd}
\newcounter{pict@widthtot}
\newcommand{\psfigaddtwo}[7]{%
\setcounter{pict@widthfst}{1*\ratio{#2+\pict@scale/2}{\pict@scale}}
\setcounter{pict@widthscd}{1*\ratio{#2+#4+\pict@scale/2}{\pict@scale}}
\setcounter{pict@widthtot}{1*\ratio{#2+#4+#6+\pict@scale/2}{\pict@scale}}
\setcounter{pict@height}{1*\ratio{#3+\pict@scale/2}{\pict@scale}}
\setlength{\unitlength}{\pict@scale}
\hbox{\hspace{-\fill}\begin{picture}(\thepict@widthtot,\thepict@height)
\put(0,0){\psfig{figure=#1,width=#2,height=#3,clip=}}
\put(\thepict@widthscd,0){\psfig{figure=#5,width=#6,height=#3,clip=}}
\SetScale{0.283466457}
\SetWidth{1.763889}
{#7}
\end{picture}}
}
\newcommand{\psfigror}[4]{%
\setcounter{pict@width}{1*\ratio{#2+\pict@scale/2}{\pict@scale}}
\setcounter{pict@height}{1*\ratio{#3+\pict@scale/2}{\pict@scale}}
\setlength{\unitlength}{\pict@scale}
\hbox{\begin{picture}(\thepict@width,\thepict@height)
\put(0,\thepict@height){\psfig{figure=#1,width=#3,height=#2,clip=,angle=270}}
\SetScale{0.283466457}
\SetWidth{1.763889}
{#4}
\end{picture}}
}
\newcommand{\psfigrol}[4]{%
\setcounter{pict@width}{1*\ratio{#2+\pict@scale/2}{\pict@scale}}
\setcounter{pict@height}{1*\ratio{#3+\pict@scale/2}{\pict@scale}}
\setlength{\unitlength}{\pict@scale}
\hbox{\begin{picture}(\thepict@width,\thepict@height)
\put(0,0){\psfig{figure=#1,width=#3,height=#2,clip=,angle=90}}
\SetScale{0.283466457}
\SetWidth{1.763889}
{#4}
\end{picture}}
}
\catcode`\@=12 
\newlength\listtextwidth

\catcode`\@=11 
\newlength{\@tabfninsert}
\newlength{\@tabfnwidth}
\newcommand{\tabfootnote}[2]{%
  \setlength{\@tabfninsert}{0.8em}
  \setlength{\@tabfnwidth}{\textwidth}
  \addtolength{\@tabfnwidth}{-\@tabfninsert}
  \addtolength{\@tabfnwidth}{-0.4em}
  \noindent\makebox[\@tabfninsert][r]{\footnotesize$^{#1}$\hfil}\hfill%
  \parbox[t]{\@tabfnwidth}{\footnotesize #2\hfill}}
\catcode`\@=12 

\def\pb1{pb$^{-1}$}

\def\kt{k_T}

\def\etjetb{E^B_{T,\rm{jet}}}
\def\etajetb{\eta^B_{\rm{jet}}}
\def\phijetb{\phi^B_{\rm{jet}}}

\def\etabr{-2 < \etajetb < 1.8}

\def\sphib{d\sigma/d\phijetb}
\def\sincphib{d\sigma/d|\phijetb|}

\def\sq2{d\sigma/dQ^2}

\def\cogam{\cos{\gamma}}
\def\cogamr{-0.7 < \cogam < 0.5}

\def\citeZEUS{{\cite{%
pl:b293:465,zeus:1993:bluebook%
}}\xspace}     
\def\citeCTD{{\cite{%
nim:a279:290,*npps:b32:181,*nim:a338:254%
}}\xspace}
\def\citeCAL{{\cite{%
nim:a309:77,*nim:a309:101,*nim:a321:356,*nim:a336:23%
}}\xspace}

\def\disent{np:b485:291}
\def\disaster{graudenz:1997,*hepph9710244}
 
\def\citeKT{{\cite{%
np:b406:187
}}\xspace}

\def\papelhighqcuad{epj:c11:427}

\def\sinistra{nim:a365:508}

\def\lepto{cpc:101:108}
 
\def\heracles{cpc:69:155,*spi:www:heracles}
 
\def\django{cpc:81:381,*spi:www:djangoh11}

\def\ariadne{cpc:71:15,*zp:c65:285}

\def\jetset{cpc:39:347,*cpc:43:367}

\def\herwig63{cpc:67:465,*jhep:0101:010,*hepph0107071}

\def\cteqfour{pr:d55:1280}

\def\mrstnininine{epj:c4:463,*epj:c14:133}

\begin{document}
\prepnum{{DESY--02--171}}

\title{
Study of the azimuthal asymmetry of jets\\ 
in neutral current deep inelastic scattering\\
at HERA
}
 
\author{ZEUS Collaboration}
\date{October, 2002}
 
\abstract{
     The azimuthal distribution of jets produced 
     in the Breit frame in high-$Q^2$ deep inelastic 
     $e^+p$ scattering has been studied with the ZEUS detector at HERA
     using an integrated luminosity of $38.6$~\pb1.
     The measured azimuthal distribution shows a structure that
     is
     well described by next-to-leading-order QCD predictions over the 
     $Q^2$ range considered, $Q^2>125$~GeV$^2$.
}
 
\makezeustitle

\def\3{\ss}                                                                                        
\pagenumbering{Roman}                                                                              
                                                   %
\begin{center}                                                                                     
{                      \Large  The ZEUS Collaboration              }                               
\end{center}                                                                                       
  S.~Chekanov,                                                                                     
  D.~Krakauer,                                                                                     
  S.~Magill,                                                                                       
  B.~Musgrave,                                                                                     
  J.~Repond,                                                                                       
  R.~Yoshida\\                                                                                     
 {\it Argonne National Laboratory, Argonne, Illinois 60439-4815}~$^{n}$                            
\par \filbreak                                                                                     
  M.C.K.~Mattingly \\                                                                              
 {\it Andrews University, Berrien Springs, Michigan 49104-0380}                                    
\par \filbreak                                                                                     
  P.~Antonioli,                                                                                    
  G.~Bari,                                                                                         
  M.~Basile,                                                                                       
  L.~Bellagamba,                                                                                   
  D.~Boscherini,                                                                                   
  A.~Bruni,                                                                                        
  G.~Bruni,                                                                                        
  G.~Cara~Romeo,                                                                                   
  L.~Cifarelli,                                                                                    
  F.~Cindolo,                                                                                      
  A.~Contin,                                                                                       
  M.~Corradi,                                                                                      
  S.~De~Pasquale,                                                                                  
  P.~Giusti,                                                                                       
  G.~Iacobucci,                                                                                    
  A.~Margotti,                                                                                     
  R.~Nania,                                                                                        
  F.~Palmonari,                                                                                    
  A.~Pesci,                                                                                        
  G.~Sartorelli,                                                                                   
  A.~Zichichi  \\                                                                                  
  {\it University and INFN Bologna, Bologna, Italy}~$^{e}$
% ...
\newpage                                                                                           
% ...
$^{\    2}$ supported by the GIF, contract I-523-13.7/97 \\                                        
% ...
                                                           %
                                                           %
\newpage   
                                                           %
                                                           %
\begin{tabular}[h]{rp{14cm}}                                                                       
$^{a}$ &  supported by the Natural Sciences and Engineering Research                               
          Council of Canada (NSERC) \\                                                             
$^{b}$ &  supported by the German Federal Ministry for Education and                               
          Research (BMBF), under contract numbers HZ1GUA 2, HZ1GUB 0, HZ1PDA 5, HZ1VFA 5\\         
$^{c}$ &  supported by the MINERVA Gesellschaft f\"ur Forschung GmbH, the                          
          Israel Science Foundation, the U.S.-Israel Binational Science                            
          Foundation and the Benozyio Center                                                       
          for High Energy Physics\\                                                                
$^{d}$ &  supported by the German-Israeli Foundation and the Israel Science                        
          Foundation\\                                                                             
$^{e}$ &  supported by the Italian National Institute for Nuclear Physics (INFN) \\                
$^{f}$ &  supported by the Japanese Ministry of Education, Science and                             
          Culture (the Monbusho) and its grants for Scientific Research\\                          
$^{g}$ &  supported by the Korean Ministry of Education and Korea Science                          
          and Engineering Foundation\\                                                             
$^{h}$ &  supported by the Netherlands Foundation for Research on Matter (FOM)\\                   
$^{i}$ &  supported by the Polish State Committee for Scientific Research,                         
          grant no. 620/E-77/SPUB-M/DESY/P-03/DZ 247/2000-2002\\                                   
$^{j}$ &  partially supported by the German Federal Ministry for Education                         
          and Research (BMBF)\\                                                                    
$^{k}$ &  supported by the Fund for Fundamental Research of Russian Ministry                       
          for Science and Edu\-cation and by the German Federal Ministry for                       
          Education and Research (BMBF)\\                                                          
$^{l}$ &  supported by the Spanish Ministry of Education and Science                               
          through funds provided by CICYT\\                                                        
$^{m}$ &  supported by the Particle Physics and Astronomy Research Council, UK\\                   
$^{n}$ &  supported by the US Department of Energy\\                                               
$^{o}$ &  supported by the US National Science Foundation\\                                        
$^{p}$ &  supported by the Polish State Committee for Scientific Research,                         
          grant no. 112/E-356/SPUB-M/DESY/P-03/DZ 301/2000-2002, 2 P03B 13922\\                    
$^{q}$ &  supported by the Polish State Committee for Scientific Research,                         
          grant no. 115/E-343/SPUB-M/DESY/P-03/DZ 121/2001-2002, 2 P03B 07022\\                    
\end{tabular}                                                                                      
                                                           %
                                                           %

\pagenumbering{arabic}
\pagestyle{plain}
\section{Introduction}
\label{sec-int}

Jet production in neutral current (NC) deep inelastic scattering (DIS) at 
high $Q^2$ (where $Q^2$ is the negative of the virtuality of the exchanged boson)
provides a testing ground for the theory of the strong interaction between quarks
and gluons, namely quantum chromodynamics (QCD). An observable of interest
is $\phijetb$, the azimuthal angle in the Breit
frame~\cite{bookfeynam:1972,*zfp:c2:237} between the lepton scattering plane,
defined by the incoming and outgoing lepton momenta, and the jets produced with
high transverse energy~($\etjetb$) in that frame.

In the Standard Model, azimuthal asymmetries arising from perturbative QCD 
effects~\mbox{\cite{prl:40:3,*pr:d18:954,*np:b144:123,*np:b145:199,*np:b148:499,*zp:c16:89,pl:b269:175,*pr:d45:46,pl:b414:205}}
are expected in the $\phijetb$ distribution. At leading order (LO), the azimuthal
dependence for unpolarised NC DIS at $Q^2\ll M^2_Z$ has the form
\begin{equation}
\label{eq:primera}
\frac{d\sigma}{d\phijetb} =  A+ B \cos \,(\phijetb) + C \cos \,(2\phijetb)\;.
\end{equation}
The current-current form of the electromagnetic interactions makes 
the cross section linear in $\cos (\phijetb)$, $\cos (2\phijetb)$, 
$\sin (\phijetb)$ and 
$\sin (2\phijetb)$. However, the coefficients of the terms in $\sin (\phijetb)$ and 
$\sin (2\phijetb)$ vanish due to time-reversal invariance and the 
absence of final-state interactions at the quark-gluon level at LO.
The coefficients $A$, $B$ and $C$ result from the convolution of the matrix elements
for the partonic processes with the parton distribution functions (PDFs) of the
proton~\cite{pl:b269:175,*pr:d45:46,pl:b414:205}. The $\cos (2\phijetb)$ term is expected from
the interference of amplitudes arising from the $+1$ and $-1$ helicity components
of the transversely polarised part of the exchanged photon, whereas the interference
between the transverse and longitudinal components gives rise to the $\cos (\phijetb)$
term. In addition, a non-perturbative contribution to the asymmetry arises from the
intrinsic transverse momentum of partons in the proton. Since such intrinsic 
transverse momenta are small~\cite{pl:b511:19}, this contribution is expected to be
negligible for jet production at high $\etjetb$~\cite{pl:b78:269}. 

Previous studies of single hadron production in NC DIS observed a $\cos \phi$ term
that was attributed to non-perturbative effects~\cite{zfp:c34:277,*pr:d48:5057}. However,
more recently, a ZEUS measurement of the azimuthal dependence of charged hadrons with
high transverse momentum in the centre-of-mass system gave evidence for perturbative
contributions to the azimuthal asymmetry~\cite{pl:b481:199}. This paper presents the
first study of the azimuthal distribution of jets with high transverse energy
in the Breit frame and the comparison with LO and 
next-to-leading-order~(NLO) QCD predictions.

\section{Experimental details}
\label{sec-exp}

These results are based on data collected in 1996-1997 with the ZEUS detector at
HERA, corresponding to an integrated luminosity of $38.6 \pm 0.6$~\pb1. The HERA
rings were operated with protons of energy $E_p=820$~GeV and positrons of energy
$E_e=27.5$~GeV. The ZEUS detector is described elsewhere~\citeZEUS. The main
components used in the present analysis are the central tracking
detector~\citeCTD, positioned in a $1.43$~T solenoidal magnetic field, and the
uranium-scintillator sampling calorimeter (CAL)~\citeCAL. The tracking detector
was used to establish an interaction vertex. The CAL covers $99.7\%$ of the
total solid angle. It is divided into three parts with a corresponding division
in the polar angle\footnote{The ZEUS coordinate system is a right-handed Cartesian
system, with the $Z$ axis pointing in the proton beam direction, referred to as
the ``forward direction'', and the $X$ axis pointing left towards the centre of
HERA. The coordinate origin is at the nominal interaction point. The
pseudorapidity is defined as $\eta=-\ln(\tan\frac{\theta}{2})$, where the polar
angle $\theta$ is taken with respect to the proton beam direction.}, $\theta$,
as viewed from the nominal interaction point: forward (FCAL,
$2.6^{\circ} < \theta < 36.7^{\circ}$), barrel (BCAL, 
$36.7^{\circ} < \theta < 129.1^{\circ}$), and rear (RCAL, 
$129.1^{\circ} < \theta < 176.2^{\circ}$). The smallest subdivision of the CAL
is called a cell. Under test-beam conditions, the CAL relative energy resolution
is $18\%/\sqrt{E {\rm (GeV)}}$ for electrons and $35\%/\sqrt{E {\rm (GeV)}}$ for
hadrons. A three-level trigger was used to select the events
online~\cite{zeus:1993:bluebook}.

As the analysis follows very closely that of the inclusive jet cross sections in
the Breit frame~\cite{incljetbf}, details about the event selection, jet finding,
systematic uncertainties and theoretical predictions are not repeated here.

The scattered-positron candidate was identified from the pattern of energy deposits
in the CAL \cite{\sinistra}. 
The kinematic region of the analysis was selected by the requirements
$Q^2>125$~GeV$^2$ and $-0.7< \cos \gamma < 0.5$, where $\gamma$ is the angle
of the scattered quark in the quark-parton
model. Cuts on this angle restrict the phase-space selection in Bjorken $x$
and the inelasticity $y$ due to the relation
\begin{equation}
 \cos \gamma = \frac{(1-y)x E_p - y E_e}{(1-y)x E_p + y E_e} \;. \nonumber
\end{equation}

The longitudinally invariant $\kt$ cluster algorithm \citeKT was used in the
inclusive mode~\cite{pr:d48:3160} to reconstruct the jets in the hadronic final
state both in data and in events simulated by Monte Carlo (MC) techniques. In data,
the algorithm was applied to the energy deposits measured in the CAL cells after
excluding those associated with the scattered-positron candidate. The jet search
was performed in the pseudorapidity ($\eta^{B}$)-azimuth ($\phi^{B}$) plane of
the Breit frame, where $\phi^B=0$ corresponds to the direction of the scattered
positron. The transverse energy of the jets in the Breit frame, $\etjetb$, was
required to be larger than 8~GeV and the pseudorapidity range was restricted to
$\etabr$. The data sample contained 8523~events, 5073~of which were one-jet,
3262~two-jet, 182~three-jet and 6~four-jet events. The $Q^2$ range covered by the
data sample extended up to $Q^2 \sim 16\; 000$~GeV$^2$; measurements of the
azimuthal distribution are presented up to a mean $Q^2$ value of $\sim 2300$~GeV$^2$.

\section{Monte Carlo studies and systematic uncertainties}

Samples of events were generated to determine the response of the detector to
jets of hadrons and to calculate the correction factors necessary to obtain the
hadron-level jet cross sections. The generated events were passed through the
GEANT~3.13-based~\cite{tech:cern-dd-ee-84-1} ZEUS detector- and trigger-simulation
programs~\cite{zeus:1993:bluebook} and were reconstructed and analysed by the same
program chain as the data. The NC DIS events were generated using the LEPTO~6.5
program~\cite{\lepto} interfaced to HERACLES~4.6.1~\cite{\heracles} via
DJANGOH~1.1~\cite{\django}. The HERACLES program includes photon and $Z$ exchanges
and first-order electroweak radiative corrections. The QCD cascade was modelled with
the ARIADNE~4.08 program~\cite{\ariadne}. The CTEQ4D~\cite{\cteqfour} parameterisations
of the proton PDFs were used. As an alternative, samples of events were generated using
the model of LEPTO based on first-order QCD matrix elements plus parton showers (MEPS).
In both cases, fragmentation into hadrons was performed using the JETSET~7.4
program~\cite{\jetset}. In these programs, the azimuthal distribution was generated
according to the LO QCD calculation.

The comparison of the reconstructed jet variables for the hadronic and the calorimetric
jets in simulated events showed that no correction was necessary for $\phijetb$ and that
the average resolution was $0.09$~radians. The sample of events generated with either
ARIADNE or LEPTO-MEPS, after applying the same offline selection as for the data, gave a
good description of the measured distributions for both the event and jet 
variables~\cite{incljetbf,thesisog}. However, a MC sample of events generated with a uniform
azimuthal distribution did not describe the observed $\phijetb$ distribution at detector
level. These comparisons establish the presence of an azimuthal modulation in the data.

The cross sections presented here were corrected to the hadron level by applying
bin-by-bin corrections to the measured distributions. The correction factors had some
dependence on $\phijetb$ due to the cuts applied to remove the effects of QED radiation
that lead to a radiated photon from the positron being misidentified as a hadronic
jet. The observed $\phijetb$ dependence of the correction factor 
was not sensitive to
the
assumed azimuthal distribution in the generator; this was confirmed by the observation
that the correction factors based on the MC sample generated with a uniform azimuthal
distribution had the same dependence on $\phijetb$. The MC programs were also 
used to evaluate the corrections for QED radiative effects, which
were negligible for the normalised cross sections.

A detailed study of the systematic uncertainties was carried out. Those that had an
effect on the shape of the azimuthal distribution were: 
\begin{itemize}
    \item
the uncertainty in the absolute energy scale of the jets; 
     \item
the uncertainty in the MC modelling of the hadronic final state, which was estimated from the differences between ARIADNE
and LEPTO-MEPS in correcting the data for detector effects;
     \item
the uncertainty in the positron identification, which was estimated
by repeating the analysis using an alternative 
technique~\cite{\papelhighqcuad} to select the scattered-positron candidate.
\end{itemize}

The relative changes in the normalised differential cross 
section induced by the variations
mentioned above
were typically smaller than the statistical uncertainties, which ranged from
$\sim 2\%$ at $Q^2 \sim 125$~GeV$^2$ up to $6\%$ at $Q^2 \sim 1000$~GeV$^2$.

\section{Perturbative QCD calculations}

The LO and NLO QCD predictions were obtained using the program DISENT \cite{\disent}.
The number of flavours was set to five and the renormalisation ($\mu_R$) and
factorisation ($\mu_F$) scales were chosen to be $\mu_R=\etjetb$ and
$\mu_F=Q$, respectively. The strong coupling constant, $\alpha_s$, was calculated
at two loops with $\Lambda^{(5)}_{\overline{MS}}=220$~MeV, corresponding to
$\alpha_s(M_Z)=0.1175$. The calculations were performed using the
MRST99~\cite{\mrstnininine} parameterisations of the proton PDFs. The results
obtained with DISENT were cross-checked by using the program DISASTER++~\cite{\disaster}.
The differences were always smaller than $1\%$. 

The perturbative QCD contribution to the terms $B$ and $C$ in Eq.~(\ref{eq:primera}) is large. At
LO in $\alpha_s$, two processes contribute to jet production in the Breit
frame: QCD-Compton scattering~(QCDC, $\gamma^*q\rightarrow qg$) and photon-gluon fusion
(PGF, $\gamma^*g\rightarrow q\overline{q}$). For the former, the scattered
gluon~(quark) preferentially appears at $\phijetb=0\:(\pi)$, whilst for the PGF
process, the $\phijetb$ dependence is dominated by the $\cos (2\phijetb)$ term and is
very similar for quarks and antiquarks. Thus, the different contributions to 
the $\cos (\phijetb)$ term from quarks and gluons tend to cancel in the $\cos (\phijetb)$
asymmetry and the predicted azimuthal distribution is very close to $A+C \cos (2\phijetb)$.
The NLO QCD correction mainly modifies the normalisation and slightly affects the shape
of this prediction. In order to test the QCD prediction for the azimuthal distribution, it
is desirable that no cut be applied to the jets in the laboratory frame; otherwise, the
azimuthal distribution can be strongly distorted by kinematic effects~\cite{pl:b414:205}.
For this reason, no such cut was used in the definition of the cross sections. 

Since the measurements refer to jets of hadrons, whereas the perturbative QCD calculations
refer to partons, the hadronisation effects were investigated by using the models of
ARIADNE, LEPTO-MEPS and HERWIG~\cite{\herwig63}. These effects were negligible~\cite{thesisog}.

\section{Results}

The cross sections presented here include every jet of hadrons in an event with
$\etjetb >8$~GeV and $\etabr$. A detailed comparison of the differential cross sections
as functions of $Q^2$, $\etjetb$ and $\etajetb$ for inclusive jet production in the
same kinematic region as used here was presented in a previous publication~\cite{incljetbf}.
At low $Q^2$ and low $\etjetb$, the NLO QCD calculations fall below the data by $\sim 10\%$.
Nonetheless, the differences between the measurements and calculations are of the same size
as the theoretical uncertainties~\cite{incljetbf}. The comparison of the shape of interest
in this publication is facilitated by normalising the predicted cross section and the
data to unity.

\begin{figure}[p]
\setlength{\unitlength}{1.0cm}
\begin{picture} (18.0,17.0)
\put (0.0,0.){\epsfig{figure=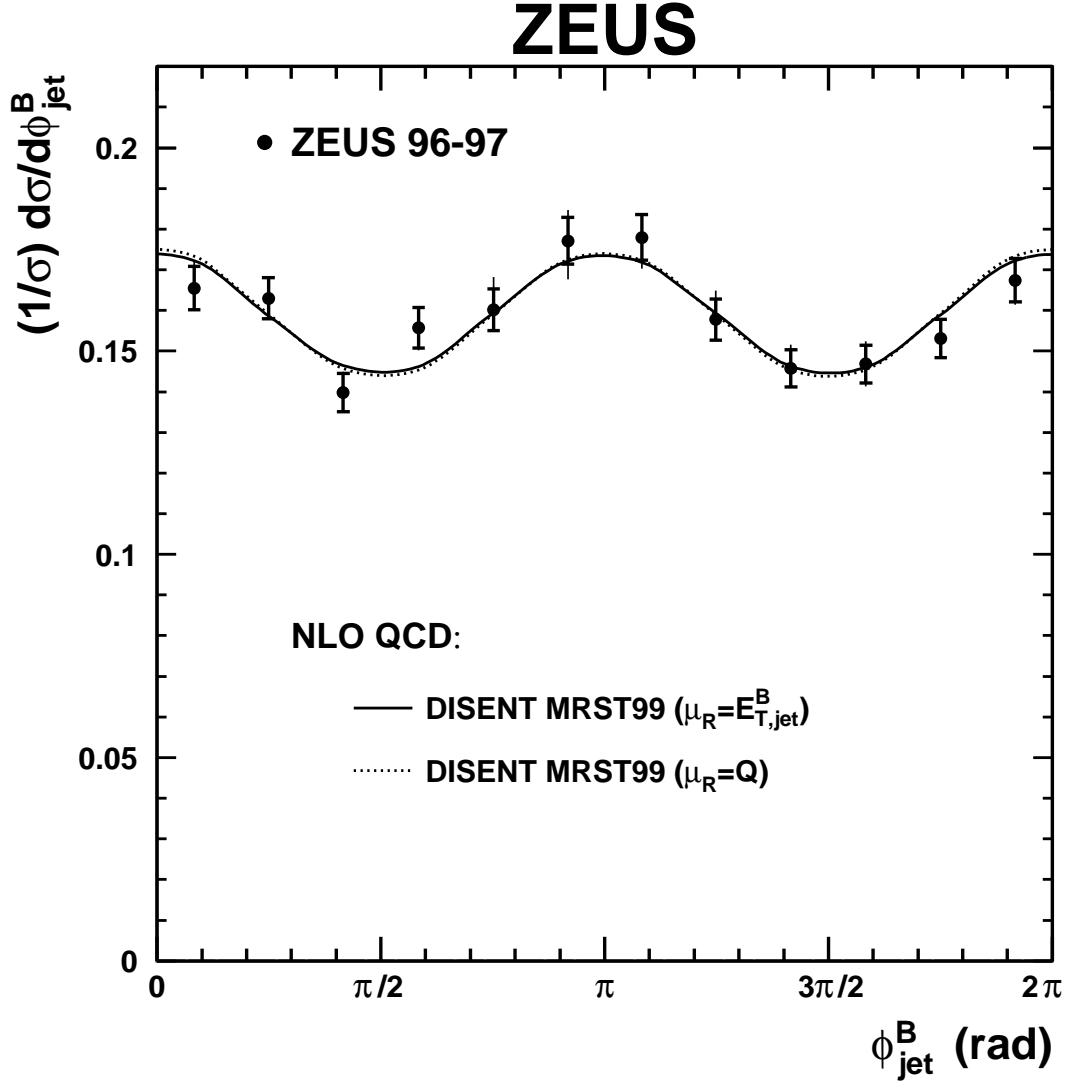,width=17cm}}
\end{picture}
\vspace{-1.0cm}
\caption{\label{fig1}
{The normalised differential cross-section $(1/\sigma)\:\sphib$ for inclusive jet production
with $\etjetb > 8$~GeV and $\etabr$ (points).
The inner error bars represent
the statistical uncertainty. The outer error bars show the statistical and
systematic uncertainties 
added in
quadrature.
The NLO QCD calculations 
using DISENT and the MRST99
parameterisations of the proton PDFs are shown for two choices of the
renormalisation scale.
}}
\end{figure}

The normalised differential cross-section $(1/\sigma)\: d\sigma/d\phijetb$ for inclusive
jet production as a function of $\phijetb$ is shown in Fig.~\ref{fig1} 
and in Table~\ref{tablaphitot}.
This distribution has
clear enhancements at $\phijetb=0$ and $\phijetb=\pi$. The NLO QCD calculations with either
$\mu_R=\etjetb$ or $Q$ reproduce the asymmetry. This comparison constitutes a precise test
of the perturbative QCD prediction for the azimuthal distribution since the theoretical
uncertainties are small. The dominant theoretical uncertainty arose from terms beyond
NLO and was estimated by varying $\mu_R$ between $\etjetb/2$ and $2\etjetb$; the effect on the
amplitude of the modulation of the distribution was $\sim \pm 1\%$. This observation complements
the ZEUS measurement of the azimuthal dependence of charged hadrons with high transverse momentum
in NC DIS~\cite{pl:b481:199}. 

\begin{figure}[p]
\setlength{\unitlength}{1.0cm}
\begin{picture} (18.0,17.0)
\put (0.0,0.){\epsfig{figure=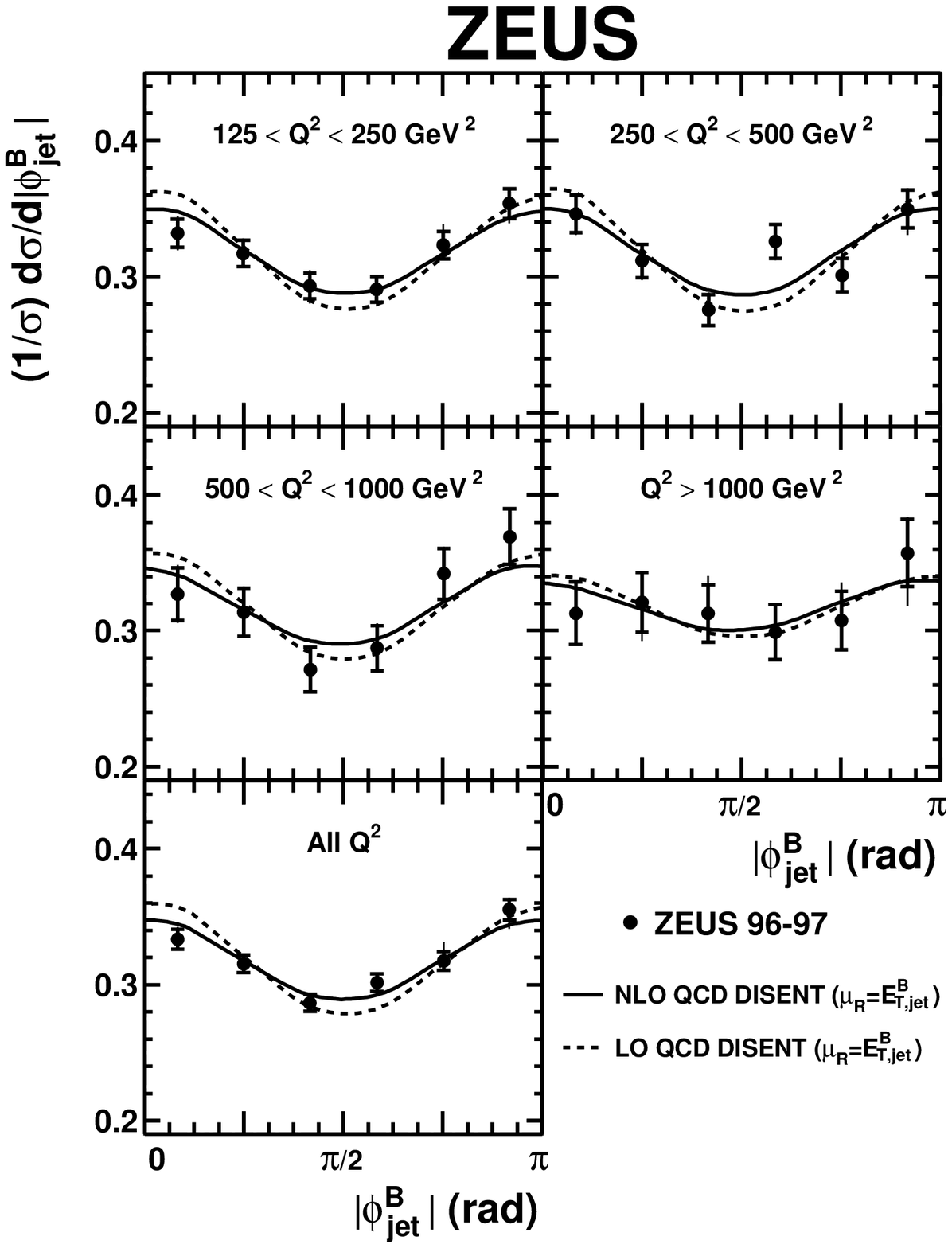,width=17cm}}
\end{picture}
\vspace{-0.50cm}
\caption{\label{fig2}
{The folded normalised differential cross-section $(1/\sigma)\:\sincphib$ for 
inclusive jet production
with $\etjetb > 8$~GeV and $\etabr$ in different 
$Q^2$ regions (points).
The inner error bars represent
the statistical uncertainty. The outer error bars show the statistical and
systematic uncertainties 
added in
quadrature.
The LO and NLO QCD calculations 
using DISENT and the MRST99
parameterisations of the proton PDFs are also shown.
}}
\end{figure}

The measurements folded about $\pi$, $|\phijetb|$,
in different regions of $Q^2$ are presented in Fig.~\ref{fig2} 
and in Table~\ref{tablaphipart}. 
The LO and NLO QCD predictions are compared to the data. The NLO
QCD prediction describes the data well, whereas the LO QCD calculations predict a larger asymmetry,
particularly in the lower $Q^2$ intervals. In both cases, the asymmetry is predicted to decrease
as $Q^2$ increases, as a result of the progressive decline of the contribution from the
PGF process.

\begin{figure}[p]
\setlength{\unitlength}{1.0cm}
\begin{picture} (18.0,17.0)
\put (0.0,0.){\epsfig{figure=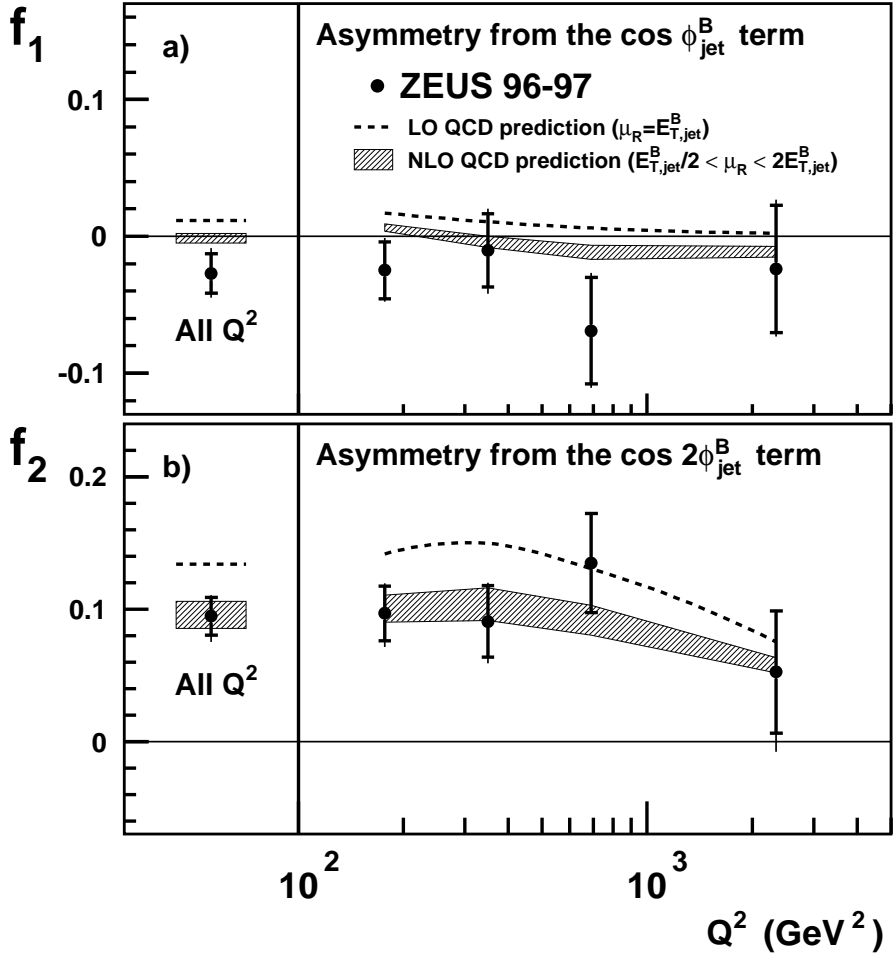,width=17cm}}
\end{picture}
\vspace{-2.0cm}
\caption{\label{fig3}
{The fitted values of a) $f_1$ and b) $f_2$ from the folded normalised
differential cross-section  $(1/\sigma)\:\sincphib$ for inclusive
jet production with $\etjetb > 8$~GeV and $\etabr$ as
a function of $Q^2$~(points). The fitted values for the entire sample,
$Q^2>125$~GeV$\:^2$, are shown on the left-hand side.
The inner error bars represent
the statistical uncertainty. The outer error bars show the statistical and
systematic uncertainties 
added in
quadrature.
The results of the fits to the LO and NLO QCD predictions using DISENT
and the MRST99
parameterisations of the proton PDFs are shown.
The shaded bands represent the 
uncertainty of the calculations due to the dependence
on the renormalisation scale. 
}}
\end{figure}

In order to perform a more quantitative study of the asymmetry and its dependence
on $Q^2$, a fit was performed to the values of $(1/\sigma)\:\sincphib$ both in the data
and in the QCD predictions. The functional form
\begin{equation}
 \frac{1}{\sigma}\:\frac{d\sigma}{d|\phijetb|} = 
 \frac{1}{\pi}\:\bigg[ 1 + f_1 \cos \,(\phijetb) + f_2 \cos \,(2\phijetb) \bigg] \nonumber
\end{equation}
was used. The parameter $f_1$~($f_2$) represents the contribution of the
$\cos \phijetb$~($\cos 2\phijetb$) term to the total asymmetry. The fitted values
of $f_1$ and $f_2$ as functions of $Q^2$ and for the entire sample with
$Q^2>125$~GeV$^2$ are shown in Fig.~\ref{fig3} and listed in Table~\ref{tablaf1f2}, together with
the LO and NLO QCD predictions and their uncertainties. 
The comparison of the LO QCD calculations for the QCDC and PGF process
shows that the asymmetry is predicted to arise mostly from the gluon-induced interactions.
The LO QCD predictions do not reproduce
the measurements. However, the uncertainty at LO is rather large. The difference between the LO
and NLO calculations has been assigned as the theoretical uncertainty of the LO predictions and
is $\sim \pm 0.04$~($\pm 0.01$) for $f_2$~($f_1$). At NLO, the dominant theoretical uncertainty
on $f_2$ ($f_1$) was that due to terms beyond NLO and was estimated by fitting the predictions
obtained with $\mu_R=\etjetb/2$ and~$2\etjetb$; it amounted 
to~$\sim \pm 0.01$ ($\pm 0.005$).
The NLO predictions for $f_1$ and $f_2$ based on calculations using $\mu_R=Q$ differed
from those using $\mu_R=\etjetb$ by as much as the estimated theoretical uncertainty.
The NLO QCD predictions are in good agreement with the measured values of $f_2$. 
For $f_1$, the observed asymmetry tends to be slightly larger and more negative than that
predicted by perturbative QCD. The measurements are consistent with the $Q^2$ dependence of $f_1$
and $f_2$ predicted by NLO QCD.

\section{Summary}    

A study of the azimuthal asymmetry for inclusive jet production in neutral current
deep inelastic $e^+p$ scattering in the Breit frame at a centre-of-mass energy of
300~GeV has been presented. Jets of hadrons were identified with the
longitudinally invariant $\kt$ cluster algorithm in the Breit frame. The normalised
cross sections as a function of the azimuthal angle of the jets in the Breit
frame are given in the kinematic region $Q^2> 125$~GeV$^2$ and $\cogamr$.
The cross sections include every jet of hadrons in the event with $\etjetb >8$~GeV
and $\etabr$.
 
The measured azimuthal distribution peaks in the directions along, and opposite to,
that of the scattered positron in the Breit frame. The NLO QCD calculations give a
good description of the observed azimuthal variation. The dependence of the azimuthal
asymmetry on $Q^2$ is also compatible with NLO QCD.

These measurements constitute a precise test of the perturbative QCD prediction for the 
azimuthal distribution since the theoretical uncertainties are small.
                                
\vspace{0.5cm}
\noindent {\Large\bf Acknowledgments}
\vspace{0.3cm}
 
We thank the DESY Directorate for their strong support and encouragement.
The remarkable achievements of the HERA machine group were essential for
the successful completion of this work and are greatly appreciated. We
are grateful for the support of the DESY computing and network services.
The design, construction and installation of the ZEUS detector have been
made possible owing to the ingenuity and effort of many people from DESY
and home institutes who are not listed as authors.   
                   
\vfill\eject

\pagestyle{plain}
  \bibliographystyle{./BiBTeX/bst/l4z_pl}
{\raggedright
\providecommand{\etal}{et al.\xspace}
\providecommand{\coll}{Collaboration}
\catcode`\@=11
\def\@bibitem#1{%
\ifmc@bstsupport
  \mc@iftail{#1}%
    {;\newline\ignorespaces}%
    {\ifmc@first\else.\fi\orig@bibitem{#1}}
  \mc@firstfalse
\else
  \mc@iftail{#1}%
    {\ignorespaces}%
    {\orig@bibitem{#1}}%
\fi}%
\catcode`\@=12

}
\vfill\eject


\begin{table}[p]
\begin{center}    
\begin{tabular}{|c|ccc|}
\hline
       $\phijetb$ {\small interval} (rad) &
       {\small $({1}/{\sigma})\;{d\sigma/d\phijetb}$} & $\;\;\;\;\;\Delta_{stat}\;\;\;$ 
       & $\;\;\;\;\;\Delta_{syst}\;\;\;$ 
   \\[.05cm]
   \hline
   \hline
   &&&\\[-0.5cm]    
      $\;\;\;\;0-\pi/6$ & $0.1655$ & {\small $\pm 0.0054$} & $^{+0.0042}_{-0.0015}$
  \\[.1cm]         
      $\pi/6-\pi/3$ & $0.1630$ & {\small $\pm 0.0051$} & $^{+0.0011}_{-0.0014}$
  \\[.1cm]         
      $\pi/3-\pi/2$ & $0.1398$ & {\small $\pm 0.0047$} & $^{+0.0020}_{-0.0008}$
  \\[.1cm]         
      $\;\;\pi/2-2\pi/3$ & $0.1557$ & {\small $\pm 0.0050$} & $^{+0.0000}_{-0.0022}$
  \\[.1cm]         
      $2\pi/3-5\pi/6$ & $0.1601$ & {\small $\pm 0.0051$} & $^{+0.0062}_{-0.0013}$
  \\[.1cm]         
      $5\pi/6-\pi\;\;\;\;\;$ & $0.1771$ & {\small $\pm 0.0057$} & $^{+0.0050}_{-0.0075}$
  \\[.1cm]
      $\;\;\;\;\;\pi-7\pi/6$ & $0.1779$ & {\small $\pm 0.0056$} & $^{+0.0016}_{-0.0052}$
  \\[.1cm]         
      $7\pi/6-4\pi/3$ & $0.1577$ & {\small $\pm 0.0050$} & $^{+0.0051}_{-0.0015}$
  \\[.1cm]         
      $4\pi/3-3\pi/2$ & $0.1458$ & {\small $\pm 0.0046$} & $^{+0.0035}_{-0.0008}$
  \\[.1cm]         
      $3\pi/2-5\pi/3$ & $0.1468$ & {\small$\pm 0.0047$} & $^{+0.0032}_{-0.0028}$
  \\[.1cm]         
      $\;\;5\pi/3-11\pi/6$ & $0.1531$ & {\small $\pm 0.0047$} & $^{+0.0017}_{-0.0014}$
  \\[.1cm]       
      $11\pi/6-2\pi\;\;\;\;\;$ & $0.1674$ & {\small$\pm 0.0054$} & $^{+0.0018}_{-0.0027}$
  \\[.1cm]
\hline
\end{tabular}
\caption{\label{tablaphitot}
Normalised differential cross-section $(1/\sigma)\:d\sigma/d\phijetb$ for
inclusive jet production with $\etjetb>8$~GeV and $-2 < \etajetb<1.8$.
The statistical and systematic
uncertainties are shown separately.
}
\end{center}
\end{table}

\begin{table}[p]
\begin{center}    
\begin{tabular}{|c|ccc||ccc|}
\hline
       \parbox{2.3cm}{\begin{center}$|\phijetb|$ {\small interval} \\ (rad)\end{center}} &
       \raisebox{.3ex}{{\footnotesize $({1}/{\sigma})\;{d\sigma/d|\phijetb|}$}} & $\Delta_{stat}$ 
       & $\Delta_{syst}$ &
       \raisebox{.3ex}{{\footnotesize $({1}/{\sigma})\;{d\sigma/d|\phijetb|}$}} & $\Delta_{stat}$ 
       & $\Delta_{syst}$ 
   \\[.05cm]
   \hline
   \hline
   &&&&&&\\[-0.5cm]    
        & \multicolumn{3}{c||}{$125 < Q^2 < 250\text{ GeV}^2$} &
          \multicolumn{3}{c|}{$250 < Q^2 < 500\text{ GeV}^2$} 
   \\[.05cm]
   \hline
   &&&&&&\\[-0.5cm]    
      $\;\;\;\;0-\pi/6$ & $0.3319$ & {\small $\pm 0.0103$} & $^{+0.0068}_{-0.0069}$ &
       $0.3461$ & {\small $\pm 0.0138$} & $^{+0.0071}_{-0.0073}$ 
  \\[.1cm]         
      $\pi/6-\pi/3$ & $0.3171$ & {\small $\pm 0.0096$} & $^{+0.0054}_{-0.0028}$ &
       $0.3116$ & {\small $\pm 0.0122$} & $^{+0.0054}_{-0.0072}$ 
  \\[.1cm]         
      $\pi/3-\pi/2$ & $0.2932$ & {\small $\pm 0.0095$} & $^{+0.0066}_{-0.0085}$ &
       $0.2754$ & {\small $\pm 0.0116$} & $^{+0.0060}_{-0.0047}$ 
  \\[.1cm]         
      $\;\;\pi/2-2\pi/3$ & $0.2907$ & {\small $\pm 0.0093$} & $^{+0.0038}_{-0.0018}$ &
       $0.3259$ & {\small $\pm 0.0126$} & $^{+0.0032}_{-0.0051}$ 
  \\[.1cm]         
      $2\pi/3-5\pi/6$ & $0.3232$ & {\small $\pm 0.0101$} & $^{+0.0120}_{-0.0021}$ &
       $0.3011$ & {\small $\pm 0.0122$} & $^{+0.0129}_{-0.0033}$ 
  \\[.1cm]  
      $5\pi/6-\pi\;\;\;\;\;\;$ & $0.3538$ & {\small $\pm 0.0109$} & $^{+0.0049}_{-0.0094}$ &
        $0.3497$ & {\small $\pm 0.0141$} & $^{+0.0059}_{-0.0126}$ 
   \\[0.05cm]
   \hline
   &&&&&&\\[-0.5cm]    
        & \multicolumn{3}{c||}{$500 < Q^2 < 1000\text{ GeV}^2$} &
          \multicolumn{3}{c|}{$Q^2 >1000\text{ GeV}^2$} 
   \\[.05cm]
   \hline
   &&&&&&\\[-0.5cm]    
      $\;\;\;\;0-\pi/6$ & $0.3268$ & {\small $\pm 0.0192$} & $^{+0.0100}_{-0.0085}$ &
       $0.3129$ & {\small $\pm 0.0229$} & $^{+0.0064}_{-0.0047}$ 
  \\[.1cm]         
      $\pi/6-\pi/3$ & $0.3136$ & {\small $\pm 0.0178$} & $^{+0.0063}_{-0.0055}$ &
       $0.3210$ & {\small $\pm 0.0220$} & $^{+0.0068}_{-0.0182}$ 
  \\[.1cm]         
      $\pi/3-\pi/2$ & $0.2713$ & {\small $\pm 0.0164$} & $^{+0.0053}_{-0.0052}$ &
       $0.3126$ & {\small $\pm 0.0211$} & $^{+0.0177}_{-0.0039}$ 
  \\[.1cm]         
      $\;\;\pi/2-2\pi/3$ & $0.2871$ & {\small $\pm 0.0167$} & $^{+0.0079}_{-0.0062}$ &
       $0.2989$ & {\small $\pm 0.0202$} & $^{+0.0027}_{-0.0009}$ 
  \\[.1cm]         
      $2\pi/3-5\pi/6$ & $0.3418$ & {\small $\pm 0.0187$} & $^{+0.0075}_{-0.0036}$ &
       $0.3074$ & {\small $\pm 0.0215$} & $^{+0.0178}_{-0.0048}$ 
  \\[.1cm]  
      $5\pi/6-\pi\;\;\;\;\;\;$ & $0.3693$ & {\small $\pm 0.0206$} & $^{+0.0062}_{-0.0120}$ &
        $0.3571$ & {\small $\pm 0.0247$} & $^{+0.0105}_{-0.0299}$ 
   \\[0.05cm]
   \hline
   &&&&&&\\[-0.5cm]  
        & \multicolumn{3}{c||}{$Q^2 > 125\text{ GeV}^2$} &
          \multicolumn{3}{c|}{} 
   \\[.05cm]
   \hline
   &&&&&&\\[-0.5cm]    
      $\;\;\;\;0-\pi/6$ & $0.3334$ & {\small $\pm 0.0072$} & $^{+0.0053}_{-0.0043}$ &
       && 
  \\[.1cm]         
      $\pi/6-\pi/3$ & $0.3153$ & {\small $\pm 0.0066$} & $^{+0.0026}_{-0.0027}$ &
       &&
  \\[.1cm]         
      $\pi/3-\pi/2$ & $0.2867$ & {\small $\pm 0.0064$} & $^{+0.0035}_{-0.0019}$ &
       &&
  \\[.1cm]         
      $\;\;\pi/2-2\pi/3$ & $0.3016$ & {\small $\pm 0.0065$} & $^{+0.0017}_{-0.0014}$ &
       &&
  \\[.1cm]         
      $2\pi/3-5\pi/6$ & $0.3176$ & {\small $\pm 0.0068$} & $^{+0.0116}_{-0.0025}$ &
       &&
  \\[.1cm]  
      $5\pi/6-\pi\;\;\;\;\;\;$ & $0.3552$ & {\small $\pm 0.0076$} & $^{+0.0044}_{-0.0117}$ &
        && 
   \\[.1cm]         
\hline 
\end{tabular}
\caption{\label{tablaphipart}
Folded normalised differential cross-section 
$(1/\sigma)\:d\sigma/d|\phijetb|$ in different regions of $Q^2$ for
inclusive jet production with $\etjetb>8$~GeV and $-2 < \etajetb<1.8$.
For details, see the caption of Table~\ref{tablaphitot}.
}  
\end{center}
\end{table}
%

%

\begin{table}[p]
\begin{center}
\mbox{
\begin{tabular}{|c|c|rcc||r|r|}
\hline                           
       & 
       $Q^2$ region (GeV$^2$) &
       &
                             $\Delta_{stat}$ & $\Delta_{syst}$ &  
      \multicolumn{1}{c|}{\parbox{2.8cm}{\footnotesize \begin{center}LO QCD\\(PGF,QCDC)\end{center}}} 
      & \multicolumn{1}{c|}{NLO QCD}
   \\[.05cm]
   \hline
   \hline
   &&&&&&\\[-0.5cm]               
      $f_1$ 
              & {\small All $Q^2$ ($Q^2 > 125$)} & $-0.0273$
                       & {\small $\pm 0.0144$} & $^{+0.0121}_{-0.0099}$ 
                                         & $0.0115$ {\small $\pm 0.0118$}
                                         & $-0.0003\; ^{+0.0025}_{-0.0044}$\\[-.1cm]
              &&&&& \multicolumn{1}{c|}{\footnotesize \hspace{1ex}($0.0236$,$-0.0013$)}&\\[.15cm]
              & {\small $125 < Q^2 < 250$} & $-0.0248$
                                      & {\small $\pm 0.0208$} & $^{+0.0113}_{-0.0093}$ 
                                      & $0.0171$ {\small $\pm 0.0100$}
                                      & $0.0071\; ^{+0.0021}_{-0.0035}$\\[-.1cm]  
              &&&&& \multicolumn{1}{c|}{\footnotesize \hspace{1ex}($0.0303$,$-0.0005$)}&\\[.05cm]
              & {\small $250 < Q^2 < 500$} & $-0.0103$
                                      & {\small $\pm 0.0268$} & $^{+0.0144}_{-0.0166}$ 
                                      & $0.0106$ {\small $\pm 0.0136$}
                                      & $-0.0030\; ^{+0.0029}_{-0.0052}$\\[-.1cm]   
              &&&&& \multicolumn{1}{c|}{\footnotesize \hspace{1ex}($0.0210$,$-0.0015$)}&\\[.05cm]
              & {\small $500 < Q^2 < 1000$} & $-0.0690$
                                      & {\small $\pm 0.0388$} & $^{+0.0166}_{-0.0150}$ 
                                      & $0.0060$ {\small $\pm 0.0161$}
                                      & $-0.0101\; ^{+0.0036}_{-0.0067}$\\[-.1cm]   
              &&&&& \multicolumn{1}{c|}{\footnotesize \hspace{1ex}($0.0152$,$-0.0029$)}&\\[.05cm]
              & {\small$Q^2 > 1000$}& $-0.0238$
                                      & {\small $\pm 0.0465$} & $^{+0.0196}_{-0.0168}$
                                      & $0.0022$ {\small $\pm 0.0122$}
                                      & $-0.0100\; ^{+0.0028}_{-0.0052}$\\[-.1cm]  
              &&&&& \multicolumn{1}{c|}{\footnotesize \hspace{1ex}($0.0089$,$-0.0009$)}&\\[.1cm]
   \hline
   \hline
   &&&&&&\\[-0.5cm]  
      $f_2$ 
              & {\small All $Q^2$ ($Q^2 > 125$)} & $0.0947$
                                         & {\small $\pm 0.0143$} & $^{+0.0068}_{-0.0133}$ 
                                         & $0.1340$ {\small $\pm 0.0356$}
                                         & $0.0984\; ^{+0.0074}_{-0.0131}$\\[-.1cm]   
              &&&&& \multicolumn{1}{c|}{\footnotesize \hspace{1ex}($0.1999$,$0.0452$)}&\\[.15cm]
              & {\small $125 < Q^2 < 250$} & $0.0969$
                                      & {\small $\pm 0.0207$} & $^{+0.0095}_{-0.0151}$ 
                                      & $0.1418$ {\small $\pm 0.0388$}
                                      & $0.1030\; ^{+0.0074}_{-0.0127}$\\[-.1cm] 
               &&&&& \multicolumn{1}{c|}{\footnotesize \hspace{1ex}($0.1880$,$0.0410$)}&\\[.05cm]
              & {\small $250 < Q^2 < 500$} & $0.0906$
                                      & {\small $\pm 0.0270$} & $^{+0.0112}_{-0.0164}$ 
                                      & $0.1496$ {\small $\pm 0.0424$}
                                      & $0.1072\; ^{+0.0088}_{-0.0158}$\\[-.1cm]  
              &&&&& \multicolumn{1}{c|}{\footnotesize \hspace{1ex}($0.2262$,$0.0632$)}&\\[.05cm]
              & {\small $500 < Q^2 < 1000$} & $0.1348$
                                      & {\small $\pm 0.0374$} & $^{+0.0044}_{-0.0082}$ 
                                      & $0.1306$ {\small $\pm 0.0356$}
                                      & $0.0950\; ^{+0.0079}_{-0.0146}$\\[-.1cm] 
                  &&&&& \multicolumn{1}{c|}{\footnotesize \hspace{1ex}($0.1982$,$0.0358$)}&\\[.05cm]
              & {\small $Q^2 > 1000$} & $0.0526$
                                      & {\small $\pm 0.0462$} & $^{+0.0086}_{-0.0387}$ 
                                      & $0.0754$ {\small $\pm 0.0160$}
                                      & $0.0594\; ^{+0.0041}_{-0.0076}$\\[-.1cm]  
                &&&&& \multicolumn{1}{c|}{\footnotesize \hspace{1ex}($0.1678$,$0.0359$)}&\\[.05cm]
\hline
\end{tabular}}
\caption{
Measured values of the parameters $f_1$ and $f_2$ in the different $Q^2$ regions.
The 
LO and NLO QCD predicted values calculated using DISENT and the MRST99 parameterisation
of the proton PDFs are shown for comparison. The quoted uncertainties in the theoretical
predictions are described in the text.
}
  \label{tablaf1f2}
\end{center}
\end{table}
%

%

%
%
\end{document}